\documentclass[notitlepage,superscriptaddress,nofootinbib,tightenlines,preprintnumbers,twocolumn]{revtex4-2}
\usepackage{amsmath,verbatim,latexsym,amssymb,indentfirst,mathrsfs,mathtools,amsthm,bbm,bm,hyperref,url,cancel,subcaption,enumitem}
\usepackage[font=small,labelfont=bf,format=plain,justification=raggedright,singlelinecheck=false]{caption}
\usepackage{verbatim,indentfirst}
\usepackage[title,titletoc]{appendix}
\usepackage{tikz}
\usetikzlibrary{arrows.meta}

\usepackage{graphicx}
\usepackage{epstopdf}
\newcommand{\caphead}[1]{{\bf #1}}

\renewcommand{\thesection}{\Roman{section}}
\renewcommand{\thesubsection}{\Roman{section} \Alph{subsection}}
\renewcommand{\thesubsubsection}{\Roman{section} \Alph{subsection} \arabic{subsubsection}}
\makeatletter
\def\p@subsection{}
\makeatother
\makeatletter
\def\p@subsubsection{}
\makeatother

\makeatletter
\newcommand\footnoteref[1]{\protected@xdef\@thefnmark{\ref{#1}}\@footnotemark}
\makeatother

\usepackage{soul}

\usepackage{CJKutf8}

\usepackage{color}
\usepackage[normalem]{ulem}

\newcommand{\A}{\mathcal{A}}
\newcommand{\B}{\mathcal{B}}

\newcommand{\sq}{{\rm sq}}

\newcommand{\Min}{ {\rm min} }   
\newcommand{\Max}{ {\rm max} }   
\newcommand{\hc}{ {\rm h.c.} }

\def\id{\mathbbm{1}}   

\newcommand{\var}{{\rm var}}  

\newcommand{\LParen}{ \bm{(} }
\newcommand{\RParen}{ \bm{)} }

\newcommand*{\Set}[1]{\left\{  #1  \right\}}

\newcommand*{\bra}[1]{\langle #1\rvert}
\newcommand*{\ket}[1]{\lvert #1 \rangle}
\newcommand*{\braket}[2]{\langle #1 \lvert #2 \rangle}
\newcommand*{\ketbra}[2]{\lvert #1 \rangle\!\langle #2 \rvert}

\begin{document}
 
\title{Quantum metrology enhanced by effective time reversal}

\author{\begin{CJK}{UTF8}{gbsn}Yu-Xin Wang (王语馨)\end{CJK}}
\affiliation{Joint Center for Quantum Information and Computer Science, NIST and University of Maryland, College Park, Maryland 20742, USA}

\author{Flavio Salvati}
\affiliation{Cavendish Laboratory, Department of Physics, University of Cambridge, Cambridge CB3 0HE, United Kingdom}

\author{David R.~M.~Arvidsson-Shukur}
\affiliation{Hitachi Cambridge Laboratory, J. J. Thomson Avenue, Cambridge CB3 0HE, United Kingdom}

\author{William F.~Braasch Jr.}
\affiliation{Joint Center for Quantum Information and Computer Science, NIST and University of Maryland, College Park, Maryland 20742, USA}

\author{Kater Murch}
\affiliation{Department of Electrical Engineering and Computer Science; University of California, Berkeley; Berkeley, CA 94720; USA.}
\affiliation{Department of Physics, University of California, Berkeley; Berkeley, CA 94720; USA}

\author{Nicole~Yunger~Halpern}
\email{nicoleyh@umd.edu}
\affiliation{Joint Center for Quantum Information and Computer Science, NIST and University of Maryland, College Park, Maryland 20742, USA}
\affiliation{Institute for Physical Science and Technology, University of Maryland, College Park, MD 20742, USA}

\date{\today}

%
%
\begin{abstract}
Quantum metrology involves the application of quantum resources to enhance measurements. Several communities have developed quantum-metrology strategies that leverage effective time reversals. These strategies, we posit, form four classes. First, \textit{mirror metrology} begins with a preparatory unitary and ends with that unitary’s time-reverse. The protocol amplifies the visibility of a small parameter to be sensed. Similarly, \textit{weak-value amplification} enhances a weak coupling's detectability. The technique exhibits counterintuitive properties captured by a retrocausal model. Using the third strategy, one simulates \textit{closed timelike curves}, worldlines that loop back on themselves in time. The fourth strategy involves \textit{indefinite causal order}, which characterises channels applied in a superposition of orderings. We review these four strategies, which we unify under the heading of \emph{time-reverse metrology}. We also outline opportunities for this toolkit in quantum metrology; quantum information science; quantum foundations; atomic, molecular, and optical physics; and solid-state physics.
\end{abstract}

{\let\newpage\relax\maketitle}


%
%
%
\section{Introduction}
\label{sec_Intro}

Quantum metrology is the study of how quantum phenomena can improve measurements. One quantum-metrology tool has enjoyed diverse applications recently: the effective reversal of time. We dub this approach \emph{time-reverse metrology}, for conciseness.\footnote{
This perspective contains several names that allude to wordy phrases, such as \emph{effective reversal of time}. In designing names, we have worked to balance practicality, descriptiveness, clarity, and convention.}
Quantum metrologists effectively reverse time in four ways:
\begin{enumerate}

   \item  We can prepare a sensor with a unitary operation $\hat{V}$, couple the sensor to a target system, and then subject the sensor to $\hat{V}^\dag$. The $\hat{V}$ and $\hat{V}^\dag$ magnify the target's impact on the sensor state. We call this strategy \emph{mirror metrology}.

   \item During \emph{weak-value amplification}, we measure the strength of a weak coupling between a probe and a target. We can probabilistically enhance the measurement precision by \emph{postselecting}, or conditioning on certain measurement outcomes. Postselection increases the precision by a factor dependent on a \emph{weak value}. This conditioned expectation value encodes information about a past event and a future event.

   \item A \emph{closed timelike curve} (CTC) is a worldline that loops back on itself in the time dimension. We can simulate CTCs by manipulating entanglement. Such a simulation can effectively teleport information backward in time. This process can effectively initialise a sensor in an optimal state, even if we do not know, at the beginning of the experiment, how best to prepare the sensor. We call this strategy \emph{time-loop metrology}.

   \item \emph{Indefinite causal order} (ICO) can benefit metrology. Channels $\A$ and $\B$ have a definite order if acting as $\A \circ \B$ or $\B \circ \A$. We can effectively apply the channels in a superposition of these orders. Such ICO enhances several metrological tasks, including parameter estimation, channel discrimination, and thermometry.
   
\end{enumerate}

Many researchers work on one class without knowing of the other advancements within that class, let alone of the other classes. A reason is that time-reverse metrology spans several disciplines: quantum metrology; quantum information science; quantum foundations; atomic, molecular, and optical physics; and solid-state physics. This perspective bridges these communities by presenting a unified view of time-reverse metrology. Table~\ref{table_compare} synopsises the relationships amongst the four classes. 

The rest of this review is organised as follows.
Section~\ref{sec_tech_intro} reviews metrological background. Sections~\ref{sec_UUdag}–\ref{sec_ICO} overview the four classes of time-reverse metrology. Section~\ref{sec_outlook} collates research opportunities across and beyond the classes.

\begin{table*}[t]
\begin{center}
\begin{tabular}{|c||c|c|c|c|} 
   \hline
    &   
   Mirror  &
   Weak-value  &
   Time-loop  &
   ICO metrology  \\
   &
   metrology  &
   amplification  &
   metrology &
   \\  \hline   \hline
   Mirror  &
   &   
   In $\leftarrow$, a unitary realises &  
   Some mirror- &
   In $\leftarrow$, we invert a \\
   metrology  &
   &
   effective time reversal. In $\uparrow$, &
   metrology experiments &
   unitary. In $\uparrow$, we \\
   &
   &
   preparation and &
   simulate CTCs.  &
   effectively implement \\
   &
   &
   measurement contribute to &
   &
   2 channels in a forward \\  
   &
   &
   apparent retrocausality. &
   &
   and an inverted order. \\ \hline
   Weak-value  &
   &   
   &
   The first time-loop- &
   Each offers robustness \\
   amplification  &
   &
   &
   metrology protocol &
   with respect to certain \\
   &
   &
   &
   enhanced weak-value &
   noise. Neither requires \\
   &
   &
   &
   amplification. &
   unitary inversion. \\ \hline 
   Time-loop  &   
   &   
   &
   &      
   In $\leftarrow$, we invert a unitary. \\
   metrology  &
   &
   &
   &
   In $\uparrow$, we invert 2 channels' \\
   &
   &
   &
   &
   order. \\  \hline
\end{tabular}
\caption{\textbf{Comparison of the four classes of time-reverse metrology.} }
\label{table_compare}
\end{center} 
\end{table*}

\section{Technical introduction}
\label{sec_tech_intro}

This section introduces notation and concepts applied throughout the perspective. We embolden vector symbols, apply boldface and hats to unit-vector symbols, and denote operators with hats. We denote a vector of Pauli operators by 
$\boldsymbol{\sigma} = \hat \sigma_x \boldsymbol {\hat{x} }+ \hat \sigma_y \hat {\boldsymbol{y}} + \hat \sigma_z \hat {\boldsymbol{z}}$.
Below, we review \emph{parameter estimation}, with which we often illustrate metrological tasks.

Often, when performing parameter estimation, we infer the parameter $\alpha \in \mathbb{R}$ in a unitary  
$\hat U_\alpha  \coloneqq  e^{-i \alpha \hat H}$. The generator $\hat H$ is a Hermitian operator. Some target system produces $\hat U_\alpha \, $; example targets include an external field, as well as a physical system that couples to the probe. We aim to infer $\alpha$ with the least error possible, given our resources.\footnote{
In some time-loop metrology (Sec.~\ref{sec_CTC}) and ICO metrology (Sec.~\ref{sec_ICO}), the inferred parameter characterises a nonunitary channel. We illustrate parameter estimation here with $\hat{U}_\alpha$ for simplicity and due to $U_\alpha$'s role in mirror metrology (Sec.~\ref{sec_UUdag}), weak-value amplification (Sec.~\ref{sec_wva}), some time-loop metrology, and some ICO metrology.}

The simplest approach to parameter estimation proceeds as follows. We prepare a probe in some state $\ket{\psi}$.\footnote{
The optimal initial state is pure. If the state were mixed, the probe would share correlations (information) with its environment. This sharing could prevent us from obtaining the maximum possible amount of information from the probe.}
The unitary evolves the probe's state to 
$\ket{\psi_\alpha}  \coloneqq  \hat U_\alpha  \ket{\psi}$.
Then, we measure the probe. A positive-operator-valued measure (POVM)
$\{ \hat M_x \}$ models the measurement~\cite{NielsenC10}. The measurement operators $\hat M_x$ are positive-semidefinite, $\hat M_x \geq 0$, and normalised as 
$\sum_x \hat M_x = \id$. The measurement yields outcome $x$ with a probability
$p_x(\alpha)  \coloneqq  \bra{\psi_\alpha} \hat M_x \ket{\psi_\alpha}$.

The \emph{Fisher information} (FI) quantifies the outcome probabilities' sensitivity to changes in $\alpha$:
\begin{align}
\label{eq_FI}
   I_\alpha  \coloneqq  \sum_x  p_x(\alpha) 
   \left[ \partial_\alpha \log p_x(\alpha) \right]^2 \, .
\end{align}
The FI bounds the precision achievable. Consider estimating $\alpha$ from $N$ trials, using any unbiased estimator $\check{\alpha}$. The estimator's variance obeys the \emph{Cr\'{a}mer–Rao bound}~\cite{46_Cramer_Mathematical,92_Rao_Information},
\begin{align}
   \label{eq_CR}
   \var \left( \check{\alpha}  \right)
   \geq  1 / \left(  N I_\alpha  \right) .
\end{align} 

Maximising the FI over all POVMs yields the \emph{quantum Fisher information} (QFI) of $\ket{\psi_\alpha}$ with respect to $\alpha$~\cite{Caves1994,Helstrom1969}:
\begin{align}
\label{eq_QFI}
   \mathcal{I}_\alpha
   \coloneqq  \max_{ \{ M_x \} } \left\{  I_\alpha  \right\}
   =  4  \left(  \braket{ \partial_\alpha \psi_\alpha}{ \partial_\alpha \psi_\alpha }
   -  \left\lvert  \braket{ \psi_\alpha }{ \partial_\alpha \psi_\alpha }  
   \right\rvert^2  \right) .
\end{align}
The QFI features in the \emph{quantum Cr\'{a}mer–Rao bound}~\cite{Helstrom1969},
\begin{align}
   \label{eq_QCR}
   \var \left( \check{\alpha}  \right)
   \geq  1 / \left(  N \mathcal{I}_\alpha  \right) .
\end{align}

To upper-bound $\mathcal{I}_\alpha$, we specify the optimal initial probe states $\ket{\psi}$. Suppose that the probe corresponds to a discrete Hilbert space, and denote $\hat H$'s extremal eigenvalues by $h_\Min$ and $h_\Max > h_\Min$. Denote corresponding eigenstates by $\ket{h_{\Min/\Max}}$.\footnote{
If $h_{\Min/\Max}$ is degenerate, $\ket{h_{\Min/\Max}}$ denotes an arbitrary corresponding eigenstate.}
The equal-weight superpositions of these states are the optimal $\ket{\psi}$s:
$( \ket{h_\Min} + e^{-i \phi} \ket{h_\Max} ) / \sqrt{2}$ \; $\forall \phi \in [0, 2 \pi)$. 
These states can achieve the maximal QFI: if $\Delta \hat H  \coloneqq  h_\Max - h_\Min$ denotes the generator's eigenvalue gap~\cite{Giovannetti_06,11_Giovannetti_Advances},
\begin{align}
\label{eq_QCR_H}
   \max_{ \ket{\psi} }  \Set{  \mathcal{I}_\alpha  }
   =  4 \max_{ \ket{\psi} }  
   \Set{  \bra{\psi}  \hat H^2  \ket{\psi}  -  \bra{\psi} \hat H  \ket{\psi}^2 }
   = ( \Delta \hat H )^2 . 
\end{align}

Returning to Eq.~\eqref{eq_QCR}, we highlight two possible scalings of the estimator's variance. Let $n$ denote (i) the number of copies of the probe or (ii) the number of applications of $\hat{U}_\alpha$ per trial. Suppose that (a) no probes are entangled and (b) every probe undergoes $\hat{U}_\alpha$ only once. At best, the estimator obeys the \emph{standard quantum limit}, 
$\var ( \check{\alpha} ) \sim 1/\sqrt{n} \, .$ Entanglement or sequential applications can enable \emph{Heisenberg scaling},
$\var ( \check{\alpha} ) \sim 1/n$~\cite{Giovannetti_06,11_Giovannetti_Advances}.

The explanations above generalise to multiparameter estimation. This task entails parameters $\alpha_j$, a matrix analogue of the QFI, and a corresponding quantum Cr\'{a}mer–Rao bound~\cite{Liu_2019, Albarelli_2020}.

\section{Mirror metrology}
\label{sec_UUdag}


Mirror metrology involves a unitary $\hat{V}$ and the time-reverse $\hat{V}^\dag$ thereof. We adapt the name from \emph{mirror benchmarking}, which features the same elements~\cite{21_Mayer_Theory}. Section~\ref{sec_UUdag_FOTOC} motivates the protocol intuitively. In Sec.~\ref{sec_UUdag_QFI}, we detail the protocol and prove that it achieves the information-bearing probe state's QFI. Section~\ref{sec_UUdag_squeeze} illustrates the protocol with examples that involve squeezing. Section~\ref{sec_UUdag_outlook} describes opportunities for future work.


%
\subsection{Intuition behind mirror metrology}
\label{sec_UUdag_FOTOC}

\begin{figure}[hbt]
\centering
\includegraphics[width=0.5\textwidth]{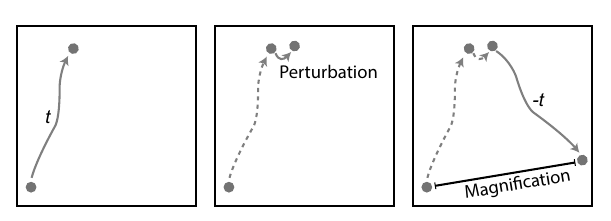}
\caption{\caphead{Motivation for mirror metrology:} Sketch of a classical chaotic system's trajectory through phase space. The system evolves forward under its Hamiltonian, undergoes a perturbation, and evolves backward under its Hamiltonian. The chaos magnifies the perturbation, rendering it easier to detect.}
\label{fig_FOTOC}
\end{figure}

The following story helps motivate mirror metrology, although without explaining its behaviour rigorously. Consider a classical chaotic system, such as a double pendulum. It undergoes the protocol sketched in Fig.~\ref{fig_FOTOC}: the system begins at some point in phase space, then evolves under its chaotic Hamiltonian. A small perturbation nudges the system's phase-space point. Then, the system evolves backward in time under the Hamiltonian. The chaos magnifies the typical such perturbation, by definition.

This protocol follows a pattern similar to that of mirror metrology. The perturbation above plays the role of the $e^{-i \alpha \hat{H} }$ introduced in Sec.~\ref{sec_tech_intro}. Forward and reverse evolutions enhance the signal's detectability. References~\cite{Rey2019,Vuletic2023,Rey2023} suggested the analogy. However, mirror metrology need not involve chaos; it relies on nontrivial, if not necessarily chaotic, evolution.

\subsection{Introduction to, and analysis of, mirror metrology}
\label{sec_UUdag_QFI}

Below, we detail mirror metrology. We then prove that it achieves the QFI of the information-bearing probe state. Finally, we review literature about mirror metrology (that does not involve squeezing, covered in Sec.~\ref{sec_UUdag_squeeze}).

Mirror metrology features the following setup and protocol. Consider a quantum probe whose Hilbert space contains a simple (unentangled, unsqueezed, etc.) state $\ket{0}$. A Hamiltonian $\hat H$ generates a unitary $ e^{-i \alpha \hat H}$ parameterised by an $\alpha \in \mathbb{R}$ of magnitude $| \alpha | \ll 1$. The following protocol implements mirror metrology:
\begin{enumerate}
    \item \label{item_prep}
    Initialise the probe in $\ket{0}$. Apply a unitary $\hat V$, preparing $| \psi   \rangle  \coloneqq   \hat V \ket{0} .$ The unitary should not commute with $\hat H$, but identifying the optimal $\hat V$ remains an open problem (Sec.~\ref{sec_UUdag_outlook}). 

    \item Evolve the probe state to
    $e ^{ - i \alpha  \hat H }| \psi \rangle 
    = e ^{ - i \alpha  \hat H }  
    \hat V \ket{0} .$

    \item \label{item_meas}
    Measure the probe in two steps. Invert the preparation unitary first, implementing 
    $\hat V^{\dag} .$
    Second, identify whether the sensor has returned to its initial state, by measuring
    $\{ \ketbra{0}{0} , \mathbb{I} - \ketbra{0}{0} \}$. Denote the $\ketbra{0}{0}$ outcome by $x{=}0$; and the alternative, by $x{=}1$.
    
\end{enumerate}

Let us compute the extractable FI and the QFI. The measurement yields the $\ket{0}$ outcome with a probability
\begin{align}
   p _{ 0 } (\alpha ) 
   & = | \langle 0 | \hat V^{\dag}
   e ^{ - i \alpha \hat  H }  
   \hat V  \ket{0} | ^{2} 
   =  | \langle 0 | 
   e ^{ - i \alpha  \hat V^{\dag} \hat H 
   \hat V }  \ket{0} | ^{2} .
\end{align}
Since $|\alpha| \ll 1$, the outcome probabilities $p _{x=0,1}  (\alpha)$ simplify to\footnote{
If $| \alpha |$ is not $\ll 1$, an adaptive strategy can achieve the QFI~\cite{95_Wiseman_Adaptive,02_Armen_Adaptive,05_Hayashi_Asymptotic,06_Fujiwara_Strong,18_Pezze_Quantum}. However, the strategy relies on a specific $V$. The main-text result governs all $V$.}
\begin{align}
   & \label{eq_PPR_P0}
   p _{ 0 } (\alpha )
   = 1 + \alpha ^{2} | \langle 0 | 
   \hat V^{\dag}   \hat H  \hat  V
   \ket{0} | ^{2} 
   \\ \nonumber & \qquad \quad \; \,
   - \alpha ^{2} | \langle 0 |  
   \hat V^{\dag}   \hat H ^{2}  \hat V   | 0 \rangle |  
   + \mathcal{O}  \left(  \alpha^{4}  \right)
   \quad \text{and} \\
   \label{eq_PPR_P1}
   & p _{ 1 } (\alpha ) 
   = 1-p _{ 0 } (\alpha ) .
\end{align}
Denote the outcome probability distribution by $\mathcal{P} (\alpha ) \coloneqq \{p _{ 0 } (\alpha ), p _{ 1 } (\alpha ) \}$. Let us substitute for the probabilities from Eqs.~\eqref{eq_PPR_P0} and \eqref{eq_PPR_P1} into the FI formula~\eqref{eq_FI}. The distribution encodes the FI
\begin{align}
   I_\alpha   \LParen  \mathcal{P} (\alpha ) 
    \RParen
   & = 4 \left( \bra{0} \hat  V^{\dag}  \hat  H^{2} \hat   V  \ket{0} 
   - \bra{0} \hat  V^{\dag}  \hat  H  \hat  V \ket{0}^{2} \right)
   \\ & \nonumber \quad \;
   + \mathcal{O}   \left(  \alpha^{2}  \right) \\
   \label{app.neq:trs.meas.cfi.gen}
   & =  4 \left( \langle \psi  | \hat H^{2}  
   |  \psi  \rangle 
   - \langle \psi  | \hat   H | \psi  \rangle^{2} \right)
   + \mathcal{O}  \left(\alpha^{2}  \right) . 
\end{align}
The final equality follows from step~\ref{item_prep} of the protocol.
This FI equals the QFI of 
$e^{-i \alpha \hat{H} } \, | \psi  \rangle$ at $\mathcal{O}(\alpha)$, by Eq.~\eqref{eq_QCR_H}. Thus, mirror metrology achieves the QFI.

Previous works state or implicitly leverage this result. Reference~\cite{Pezze2016} synopsises the argument above. In~\cite{Jing2024}, a similar argument focusses on certain $\hat V$s and $\hat H$s. Other references address general mirror metrology, arguing that it achieves the QFI of $e^{-i \alpha \hat H} \hat V \ket{0}$~\cite{Oberthaler2016,Zhou2024}. The review~\cite{Davidovich2022} illustrates mirror metrology with examples. Researchers have also extended mirror metrology to include ancilla systems~\cite{Jing2024}.

Experiments have realised mirror metrology. The protocol has been performed (in the absence of squeezing) with trapped ions~\cite{Wineland2004,Wineland2005,Leibfried2019}, a Rydberg atom and an oscillator~\cite{Dotsenko2016}, and nitrogen-vacancy centres in diamond~\cite{deLeon2025}. Many optical interferometers implement mirror metrology, casting $V$ as a beamsplitter unitary~\cite{Caves1981,White2007}.

%
\subsection{Mirror metrology based on squeezing}
\label{sec_UUdag_squeeze}

In much mirror metrology, the preparation involves squeezing. Squeezing is an operation that reduces one observable's uncertainty at the expense of another observable's~\cite{Holevo2011}. We illustrate squeezing-based mirror metrology with single-mode squeezing, spin squeezing, and two-mode squeezing.

First, we motivate single-mode mirror metrology. Consider a bosonic mode -- equivalently, a harmonic oscillator -- associated to a position operator $\hat{Q}$ and a momentum operator $\hat{P}$. Suppose that a field imprints a parameter $\alpha$ on the oscillator's state at a rate $g$, through the Hamiltonian
$\hat  H = g \alpha \hat{Q}$. This Hamiltonian kicks the oscillator's momentum:
$e^{i\hat H t} \, \hat{P} \, e^{-i\hat H t} 
= \hat{P} - g \alpha t \hat \id$. 
The momentum's displacement reveals $\alpha$, if we know $g$. Suppose we have prepared the oscillator in a state $\ket{\psi}$. The momentum displacement's expectation value forms an estimator:
\begin{align}
   \check{\alpha}
   = \left( \bra{\psi} \hat{P} \ket{\psi}
   - \bra{\psi}  e^{i\hat H t} \, \hat{P} \, e^{-i\hat H t} \ket{\psi} \right)
   / (gt) .
\end{align}
The optimal $\ket{\psi}$, a $\hat{P}$ eigenstate, achieves the maximal QFI~\cite{Holevo2011}. Momentum eigenstates are unphysical, but squeezed states can approximate them.

We parameterise single-mode squeezed states as follows~\cite{CVRMP2012}. Define 
$\hat a \coloneqq (\hat Q + i \hat P)/\sqrt{2}$ as the mode's lowering operator and $\ket{0}$ as the vacuum state. Define
$\hat Q_\varphi 
\coloneqq \hat{Q} \, \cos (\varphi) 
+ \hat{P} \, \sin (\varphi)$ as the quadrature operator associated to the phase-space vector tilted upward through an angle $\varphi$ from the $x$-axis.
The squeezing Hamiltonian
$\hat H_\sq (\varphi) 
  \coloneqq \frac{i}{2} \, ( e^{-2i \varphi} \, \hat a^2 - \hc )$ 
can squeeze $\hat Q_\varphi $. 
This Hamiltonian generates a unitary that maps the vacuum to the squeezed state
$\ket{r, \varphi}
 \coloneqq  e^{-ir \hat H_\sq (\varphi) }  \,  \ket{0}$.
Suppose that the angle $\varphi = \pi/2$. As the squeezing parameter $r \to \infty$, the state $\ket{r, \varphi}$ increasingly resembles a momentum eigenstate.

Squeezing facilitates mirror metrology for sensing $\alpha$. The probe begins in $\ket{0}$, undergoes squeezing by $e^{-ir \hat H_\sq (\varphi) } $, and evolves under $\hat H$. We then antisqueeze the sensor (evolve it under $e^{ir \hat H_\sq (\varphi) } $). Finally, we measure 
$\{ \ketbra{0}{0} , \id - \ketbra{0}{0} \}$.
The unitaries condense into\footnote{
We apply the identity
$e^{ir \hat H_\sq (\varphi {=} \pi/2) }  \,
\hat{Q}  \,
e^{-ir \hat H_\sq ( \varphi {=} \pi/2) }
=  e^r  \,  \hat{Q}$. }
\begin{align}
\label{eq:paramp.evol}
   e^{ir \hat H_\sq (\varphi {=} \pi/2 ) }  \,
   e^{-i g \alpha \hat{Q} t }  \,
   e^{-ir  \hat H_\sq  ( \varphi {=}  \pi / 2) }
   = e^{-i e^r   g \alpha \hat{Q} t }  \, .
\end{align}
The protocol effectively boosts the signal–sensor coupling strength, $g$, by a factor $e^r$. Hence the protocol effects \emph{parametric amplification}~\cite{Caves1982}. Trapped-ion experiments have demonstrated this principle~\cite{Allcock2019}.

Not only a bosonic mode, but also a collection of spins, can undergo squeezing.\footnote{
The spins are typically qubits. However, we use the term \emph{spins} for generality and for consistency with the literature.}
Various Hamiltonians enact spin squeezing, as discussed in App.~\ref{sec:app.spin.squeezing}.
Examples include the one-axis-twisting~\cite{Ueda1993}, the two-axis-twisting~\cite{Ueda1993}, the Lipkin-Meshkov-Glick~\cite{LMG1965}, and general spin–spin interacting~\cite{Yao2024} Hamiltonians. Mirror metrology offers two benefits when implemented with spin squeezing. First, an estimator's variance can scale in Heisenberg fashion with the number of spins~\cite{SchleierSmith2016}. Second, the strategy offers robustness with respect to readout error~\cite{SchleierSmith2016,Haine2017,Clerk2023}. Neutral-atom experiments have showcased these advantages~\cite{Kasevich2016,Vuletic2022,Vuletic2023}.
 
An early squeezing-based mirror-metrology proposal involves two-mode squeezing~\cite{Klauder1986}. The protocol features two bosonic modes, associated to annihilation operators $\hat a$ and $\hat b$. Suppose that mode $a$ can undergo the phase shift   
$e^{i \alpha \hat a^\dag \hat a}$.
To sense $\alpha$, we prepare both modes in their vacuum states: $\ket{0,0}$. The unitary 
$\hat S  (r) \coloneqq e ^{r (\hat{a}^{\dagger} \hat{b}^{\dagger} - \hat{b} \hat{a})}$
introduces two-mode squeezing.
Mode $a$ undergoes the phase shift. The unitary 
$\hat S^{-1}(r)$ unsqueezes the modes. Finally, we measure each mode's photon number. 
This protocol achieves the QFI of
$e^{i \alpha \hat a^\dag \hat a }  \,  \hat S(r) \ket{0,0}$. 

Two features of the two-mode protocol merit mentions. First, the protocol resembles standard optical interferometry~\cite{Caves1981}. However, $\hat V$ (step~\ref{item_prep} in Sec.~\ref{sec_UUdag_QFI}) manifests here as a two-mode-squeezing unitary, rather than as a beamsplitter. Second, the two-mode-squeezing unitaries form the SU(1,1) symmetry group. Therefore, the two-mode protocol above is called \emph{SU(1,1) interferometry}~\cite{Klauder1986}.

Experimentalists have implemented the original SU(1,1) interferometer and variations thereon. Platforms used include photonics~\cite{ZhangWeiping2011,ZhangWeiping2014,Lett2017}, atoms~\cite{Oberthaler2016}, and ions~\cite{Bollinger2021}. Applications have been proposed for the Laser Interferometer Gravitational-Wave Observatory (LIGO)~\cite{Agarwal2010}, for sensing the strength of photon-number decay~\cite{Bash2023,Davidovich2024absorption}, and for imaging~\cite{Guha2025}.

\subsection{Outlook for mirror metrology}
\label{sec_UUdag_outlook}

The mirror-metrology template offers opportunities in and beyond the parameter estimation of Sec.~\ref{sec_tech_intro}. Below, we present open questions about implementations. We then summarise possible applications to quantum information processing outside parameter estimation. Section~\ref{sec_CTC} details how certain mirror-metrology experiments simulate CTCs.

The mirror-metrology protocol begins with a unitary $\hat V$ (step~\ref{item_prep} in Sec.~\ref{sec_UUdag_QFI}). What, generally, is the optimal form of $\hat V$? Consider any $\hat V$ that prepares an equal-weight superposition 
$( \ket{h_\Min} + e^{-i \phi} \ket{h_\Max} ) / \sqrt{2} \, ,$ wherein $\phi \in [0, 2\pi)$. Every such $\hat V$ can provide the QFI, maximised over $\hat{V}'$, of 
$e ^{ - i \alpha  \hat H } \hat{V}' \ket{0}$ [Eq.~\eqref{eq_QCR_H}]. Yet implementing such $\hat V$s may pose challenges; for example, they may cost many computational gates. Furthermore, noise can affect the optimal $\hat{V}$. Ideally, $\hat V$ optimises the QFI and hardware efficiency. The compilation of such $\hat V$s is undergoing active research~\cite{SchleierSmith2016,Jing2024}. 

A related opportunity concerns decoherence. Consider performing mirror metrology using spins subject to spontaneous decay. The protocol may achieve the decayed probe state's QFI~\cite{Kielinski2024}. Mirror metrology merits extensions to sensors subject to general decoherence~\cite{Lukin2025}. Perhaps we can gain insights from weak-value amplification and ICO metrology, which offer robustness with respect to certain noise (Sections~\ref{sec_wva} and~\ref{sec_ICO}).

Beyond the parameter estimation of Sec.~\ref{sec_tech_intro}, the mirror-metrology template can facilitate further quantum information processing. The proof in Sec.~\ref{sec_UUdag_QFI} generalises to quantum state discrimination and noise estimation~\cite{Davidovich2024absorption}; the \emph{Dolinar receiver} illustrates how~\cite{Dolinar1973}. Such a receiver optimally distinguishes two coherent states by implementing a displace-probe-replace protocol. Also, prepare-evolve-reverse sequences enable Hamiltonian amplification~\cite{Arenz2020,Burd2024,Arenz2024} and the detection of higher-order out-of-time-order correlators~\cite{GoogleOTOC2025,ZhangOTOC2005}. More extensions might await discovery.

\section{Weak-value amplification}
\label{sec_wva}

\begin{figure}[t]
\centering
\includegraphics[width=0.5\textwidth]{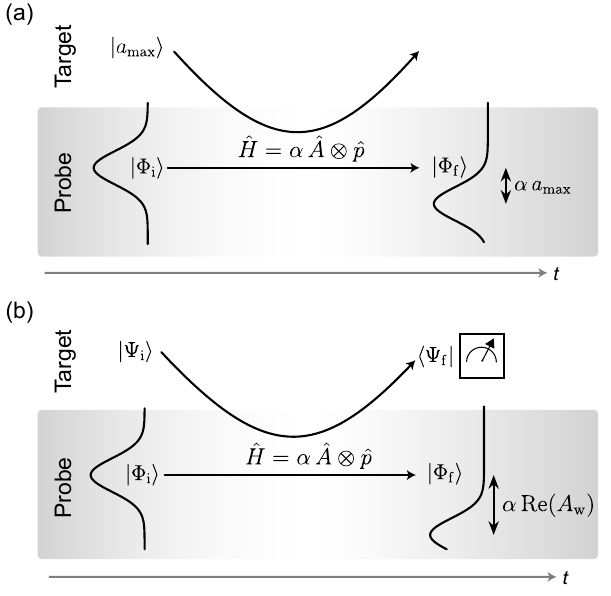}
\caption{\textbf{Weak-value amplification:} 
In each diagram, time progresses from left to right. The topmost curve represents the target; and the black horizontal line, the probe. The target is prepared in $\ket{\Psi_{\mathrm{i}}}$; and the probe, in $\ket{\Phi_{\mathrm{i}}}$. A weak interaction imprints information about the target observable $\hat{A}$ onto the distribution over the probe's possible positions. We measure the probe's position at the end of the experiment.
(a) We do not postselect the target before measuring the probe. Preparing the target in $\ket{a_{\max}}$ displaces the probe's position distribution through the greatest possible amount, $\alpha\,a_{\max} \, .$
(b) Before measuring the probe, we postselect the target on $\ket{\Psi_{\mathrm{f}}}$. 
The probe's position distribution is displaced through $\alpha\, \mathrm{Re}(A_{\mathrm{w}})$. If 
$\alpha\, \mathrm{Re}(A_{\mathrm{w}}) 
> \alpha \, a_{\max} \, ,$ weak-value amplification results.
}
\label{Fig_WVA}
\end{figure}

Metrology hinges on useful input states. Consider measuring an unknown interaction’s strength. Commonly, one optimises the system's state just before the interaction. In certain experiments, one can boost the achievable sensitivity by optimising also the postinteraction state. \emph{Weak-value amplification} exemplifies this opportunity~\cite{Zhang2015Precision, Alves2015, Dressel2012, Kofman2012, Starling2010Frequency, Dixon2009BeamDeflection, Hosten2008Science, Ritchie1991}. Below, we describe how postselecting output states effectively influences previous quantum interactions. This apparent retroaction can increase the FI attainable per measured probe. 

Weak measurements allow one to probe a target quantum system without disturbing it significantly \cite{jordan2024quantum}; von Neumann's measurement model describes the process \cite{jordan2024quantum, vonNeumann1932}. Consider a target associated to a discrete Hilbert space. This target will entangle with the probe, unlike the classical-field target in mirror metrology. Denote by $\hat{A}=\sum_n a_n \ketbra{a_n}{a_n}$ any observable defined on the target's Hilbert space. Let $a_\Min$ denote the least eigenvalue; and $a_\Max$, the greatest. 

The probe is a continuous-variable system. Let  $\hat P$ and $\hat Q$ denote its natural dimensionless position and momentum observables: $[\hat Q, \hat P] = i$. Denote by $\ket{q}$ the position eigenstate associated with the eigenvalue $q$.

A common weak measurement proceeds as follows. We prepare the target in $\ket{\Psi_{\mathrm{i}}}\coloneqq\sum_n c_n \ket{a_n}$, wherein $\sum_n |c_n|^2=1$, and the probe in $\ket{\Phi_{\mathrm{i}}} \coloneqq \int dq \: \phi(q) \ket{q}$. The wave function $\phi(q)$ is real, is centred at $q=0$, and has a standard deviation $\Delta\phi$. The target's $\hat{A}$ observable couples weakly to the probe's momentum with a strength $\alpha \gtrsim 0$. The unitary $\hat{U}_{\alpha} \coloneqq \exp(-i \alpha \, \hat{A} \otimes \hat{P})$ encodes information about the target in the probe state:
\begin{equation}
 \hat{U}_{\alpha}
\ket{\Psi_{\mathrm{i}}}\ket{\Phi_{\mathrm{i}}} =  \sum_n  \int dq \: c_n \, \phi(q - \alpha \, a_n)  \ket{a_n} \ket{q} . 
\end{equation}
This $\hat{U}_\alpha$ differs from that in Sec.~\ref{sec_tech_intro} (and so from the $\hat{U}_\alpha$ in mirror metrology and some time-loop metrology) by acting on the target-and-probe composite, rather than on the probe alone. 
We then measure the probe's position, $\hat Q$, projectively. If the probe begins with a wide wave function, $\Delta \phi \gg \alpha (a_{\max} - a_{\min})$, the interaction perturbs the joint state weakly. Across many trials, we infer the probability distribution over the possible final probe positions. The distribution peaks at $\alpha \langle \hat{A} \rangle \equiv \alpha \bra{\Psi_{\mathrm{i}}}\hat{A}\ket{\Psi_{\mathrm{i}}}$.

A common task is to measure the weak coupling strength, $\alpha$. To achieve maximum probe sensitivity in the setting above, one should initialise the target in the $\hat{A}$ eigenstate associated to the greatest eigenvalue: $\ket{ \Psi_{\mathrm{i}} } = \ket{a_{\max}}$.  This choice leads to the largest separation between the peaks of the output distribution, $ \phi(q - \alpha \, a_{\max})$, and corresponding input distribution, $\phi(q)$ (Fig. \ref{Fig_WVA}a). 

Suppose that, before measuring the probe, we measure the target and postselect on the outcome associated to some state $\ket{\Psi_{\mathrm{f}}}$. What form does the final probe state $\ket{\Phi_{\mathrm{f}}}$ have? (The target may begin in any state $\ket{\Psi_{\mathrm{i}}}$, not necessarily in $\ket{a_{\rm max}}$, in this protocol.) The answer depends on the \emph{weak value} of $\hat{A}$ \cite{jordan2024quantum, Aharonov2008,  Duck1989, Aharonov1988},
\begin{equation}
\label{Eq:wv}
    A_{\mathrm{w}} 
    \coloneqq \frac{\bra{\Psi_{\mathrm{f}}} \hat A \ket{\Psi_{\mathrm{i}}}}{\langle \Psi_{\mathrm{f}} | \Psi_{\mathrm{i}} \rangle} \, .
\end{equation}
We can express $\ket{\Phi_{\mathrm{f}}}$ in terms of $A_{\mathrm{w}}$ under two weak-measurement conditions. First, the probe's initial wave function is broad:
$\Delta \phi \gg \alpha\,|A_{\mathrm{w}}|$. 
Second, consider Taylor-approximating
$\bra{\Psi_{\mathrm f}}e^{-i\alpha \hat A\otimes \hat P}\ket{\Psi_{\mathrm i}}$
in $\alpha$. The interaction is weak enough that a linear approximation is highly accurate~\cite[Eqs.~(20)--(21)]{Duck1989}. Under these conditions, the protocol transforms the probe's state partially via the translation operator
$e^{-i\alpha {\rm Re}(A_{\mathrm{w}}) \hat P}$ \cite{Duck1989}. This transformation shifts the probe's position-basis wave function by $ \alpha \, \mathrm{Re}\left( A_{\mathrm{w}} \right)$ (Fig. \ref{Fig_WVA}b):
\begin{align}
   \ket{\Phi_{\mathrm{f}}}
   & \approx C \braket{ \Psi_{\rm f} }{ \Psi_{\rm i} }
   e^{\alpha {\rm Im} ( A_{\rm w} ) \hat{P} }
   \int dq \,
   \phi \LParen q - \alpha {\rm Re} ( A_{\rm w} ) \RParen
   \ket{q} . 
\label{eq:weak}
\end{align}
(App.~\ref{sec:app.weak.value.expansion}). $C$ denotes a normalisation factor. $\mathrm{Re}\left(A_{\mathrm{w}} \right)$ can lie outside $\hat{A}$'s spectrum, enabling  weak-value amplification, as described in the next paragraph.

The postselection can improve the inference of the weak-coupling parameter $\alpha$. To see how, recall the penultimate paragraph's postselection-free strategy. Upon concluding with an optimal probe measurement, the strategy yields the FI
\cite{Pang2015, Pang2014EntanglementWVA}
\begin{equation}
   I_{\alpha} 
   \approx \big\langle \hat{A}^2 \big\rangle 
   / \left(\Delta \phi \right) ^2  .
\label{eq:Fish}
\end{equation}
Because $\Delta \phi$ is large, we can glean little information about $\alpha$ per measured probe. Therefore, viable signal-to-noise ratios often require high probe-state intensities. However, high intensities create problems: a detector must operate below its saturation intensity \cite{Harris2017}. Moreover, after a detector registers a probe, a time lag follows before the detector can register again \cite{Uzunova2022, Lita2022, Valivarthi_2014, Dixon2009UltrashortDeadTime}. 
We can mitigate these obstacles by measuring the target and, only upon obtaining the $\ket{\Psi_{\mathrm{f}}}$ outcome, measuring the probe. We can thereby obtain the \emph{postselected FI} \cite{Pang2015},
\begin{equation}
   I_{\alpha}^{\mathrm{ps}} 
   \approx |A_{\mathrm{w}}|^2
   / \left( \Delta \phi \right) ^2  .
\label{eq:PSFish}
\end{equation}
Certain states $\ket{\Psi_{\mathrm{f}}}$ increase $|A_{\mathrm{w}}|^2$ above $\langle \hat{A}^2 \rangle$. Thus, $I_{\alpha}^{\mathrm{ps}}$ can exceed $I_{\alpha}$. Across trials, the postselection discards information \cite{Pang2014EntanglementWVA}. However, we can concentrate the FI in the postselected-on events \cite{Liu2022, Li2017, Pang2015}.

Weak-value amplification has become a widely used metrological tool. The technique gained prominence after enabling the first observation of the spin Hall effect of light (a polarisation-dependent displacement of photons at an interface) \cite{Hosten2008Science}. Weak-value amplification transduces hard-to-detect signals into accessible readouts under realistic experimental conditions -- in the presence of technical noise and detector constraints~\cite{Zhu2025WVAReview, Xu2024WVAReview, Jordan2014PRX}. Examples of such imperfections include saturation, digitisation, and time-correlated readout noise. This robustness to certain noise likens weak-value amplification to ICO metrology (Sec.~\ref{sec_ICO}).

Experiments illustrate the advantage offered by weak-value amplification; we illustrate with two recent ones. Jiang \emph{et al.} \cite{Jiang2025} applied weak-value amplification to Rydberg-atom sensing. They mapped a small microwave-induced phase shift to a large, measurable displacement of an optical probe. Their readout achieved a sensitivity gain of 5 to 6~dB over a state-of-the-art superheterodyne scheme \cite{Jing2020}. In principle, this method can approach the atomic shot-noise limit whilst mitigating technical noise.

Huang \emph{et al.} combined weak-value amplification with double-slit interferometry~\cite{Huang2025}. They weakly coupled an interarm time delay 
to a beam's transverse spatial mode. Postselection generated a large real weak value. Thus, the authors amplified a small temporal delay $\tau$ into a spatial fringe displacement $A_{\mathrm{w}}\tau$. They resolved few-attosecond delays using only narrow-band light and standard imaging optics. Compared with conventional interferometry \cite{Schweinberger2019}, their scheme improved the signal-to-noise ratio by up to two orders of magnitude.

Other applications of weak-value amplification include measurements of small beam deflections \cite{Gorodetski2012Chirality, Park2012WeakValue, Brunner2010PhaseShifts, Dixon2009BeamDeflection}, phases \cite{Song2021InverseWVA, Xu2013PhaseEstimation}, thin-film thicknesses \cite{Zhou2012SpinHall, Kong2012SpinHallBrewster, Ren2012SpinHallMagnetic}, velocities \cite{Viza2013Velocity}, frequency shifts \cite{Starling2010Frequency}, temperature changes \cite{SalazarSerrano2015FBG, Egan2012Thermostat}, and optical roll angles \cite{Gillmer2018VibrationSuppression}. Weak-value amplification has been implemented with tabletop optics \cite{Hallaji2017SinglePhoton, Pfeifer2011OpticalActivity, Ritchie1991}, integrated optical waveguides \cite{Song2021InverseWVA}, matter-wave interferometers \cite{Denkmayr2014QuantumCheshireCat}, solid-state systems \cite{Zilberberg2011}, superconducting circuits \cite{Monroe2021WeakMeasurement}, trapped ions \cite{Pan2020WeakToStrong}, and atomic-vapour cells \cite{Qu2020SubHertzResonance}. 

Having surveyed weak-value amplification's benefits, we describe its apparent retrocausality. According to Eqs.~\eqref{eq:weak} and~\eqref{Eq:wv}, the probe's final state ($\ket{\Phi_{\rm f}}$) depends on the target’s postselected state ($\ket{\Psi_{\rm f}}$). Yet the postselection occurs \emph{after} target–probe interaction. Hence the target's postselection appears to influence the probe retrocausally. This conclusion may seem unsurprising:
we can run any experiment with input states drawn from some distribution, then cherry-pick the outcomes. Yet in classical experiments, in which operations commute, postselection cannot outperform an optimal input state \cite{Arvidsson-Shukur2020}. This restriction does not extend to weak-value amplification, in which $I_{\alpha}^{\mathrm{ps}}$ can exceed $\max_{\ket{\Psi_{\mathrm{i}}}} \{ I_{\alpha} \}$. This advantage follows from contextuality \cite{Thio2024, Arvidsson-Shukur2020, Kunjwal19, Pusey2014}, a genuinely nonclassical phenomenon \cite{Spekkens05}.

The target's initial state ($\ket{\Psi_{\mathrm{i}}}$) and final state ($\ket{\Psi_{\mathrm{f}}}$) influence the probe equally [Eqs.  \eqref{eq:weak} and \eqref{Eq:wv}]. This time-symmetric influence motivates a reinterpretation of time in quantum mechanics. The two-state-vector formalism provides one such interpretation \cite{Aharonov2008, Aharonov1988, Aharonov1964}. Consider preparing a system in $\ket{\Psi_{\mathrm{i}}}$ at a time $t_1$, then postselecting the system into $\ket{\Psi_{\mathrm{f}}}$ at a time $t_2>t_1$. 
Let the times $t,t'\in[t_1,t_2]$, and denote the time-ordering operator by $\mathcal{T}$. A unitary $W(t',t)$ evolves the state from $t$ to $t'$:
\begin{equation}
    W(t',t) \coloneqq \mathcal{T}
    \exp\!\left( -i\int_{t}^{t'} ds \, \hat H(s)\right) .
\end{equation}
The two-state-vector formalism describes the system with a forward-evolving state 
$\ket{\Psi_{\mathrm{i}} (t)} 
\coloneqq W (t, t_1)\ket{\Psi_{\mathrm{i}}}$ and a backward-evolving state 
$\bra{\Psi_{\mathrm{f}} (t)} 
\coloneqq \bra{\Psi_{\mathrm{f}}} W^{\dagger}(t, t_2)$.
They form the \emph{two-state vector} 
$\langle \Psi_{\mathrm{f}} (t) | \,  |\Psi_{\mathrm{i}} (t) \rangle $. Within this framework, the time-$t$ weak value $A_{\rm w}$ is a time-symmetric expectation value conditioned on the system's past and future: in a generalisation of Eq.~\eqref{Eq:wv},
\begin{equation}
\label{Eq:wv_tsvf}
    A_{\mathrm{w}} = \frac{\bra{\Psi_{\mathrm{f}} (t)} \hat A \ket{\Psi_{\mathrm{i}}(t)}}{\langle \Psi_{\mathrm{f}} (t)| \Psi_{\mathrm{i}} (t)\rangle} \, .
\end{equation}
[In Eq.~\eqref{Eq:wv}, $W(t, t_1) = W(t, t_2) = \id$.]
The two-state-vector formalism extends Bayesian probability theory to quantum experiments \cite{Hofmann2012, Johansen2007, Steinberg1995}. 

Postselection's benefits extend beyond weak-value amplification. Generalised postselective metrology can distil information about $L$ parameters $\bm{\alpha}=(\alpha_1,\alpha_2,\ldots,\alpha_L)$ from $n$ states $\hat{\rho}_{\bm{\alpha}}$ into $m \ll n$ states \cite{ArvShu24, Jenne2021, Arvidsson-Shukur2020}. If $\hat{\rho}_{\bm{\alpha}}$ is pure, we can distil the information optimally, using a known protocol: we can make $m/n$ arbitrarily small without losing information \cite{Jenne2021}. In experiments, however, $\hat{\rho}_{\bm{\alpha}}$ decoheres. Consequently, the procedure loses information, and no general optimal strategy is known \cite{Salvati2024}. To improve information distillation in the presence of decoherence, we might apply (i) analogies with ICO metrology, which offers resilience to certain decoherence (Sec.~\ref{sec_ICO}), or (ii) the equivalences between quantum experiments and retrocausal effects. Item (ii) may also enable the application of postselection to metrology beyond parameter estimation, such as channel discrimination.

\section{Time-loop metrology}
\label{sec_CTC}

A CTC is the trajectory of a hypothetical particle that alternates between moving forward and backward in time \cite{Morris_1988}. General relativity allows for CTCs, although many physicists expect them not to exist \cite{Godel_1949,Hawking_1992}. Quantum mechanics can accommodate CTCs in multiple ways. For instance, suppose a particle correlates with another system before travelling to the past. We can posit that the correlations are identical when the particle returns to the future~\cite{Lloyd11-2,11_Lloyd_Closed}. Henceforth, we refer only to CTCs of this type. We can simulate such CTCs by manipulating entanglement and postselecting on certain measurement outcomes~\cite{11_Lloyd_Closed}: the same mathematics describes these quantum-circuit protocols and CTCs. This equivalence suggests that entanglement manipulation can enable metrological advantages that would arise from sending information backward in time. This insight has motivated time-loop quantum metrology.

\begin{figure}
    \centering
\includegraphics[width=0.5\textwidth]{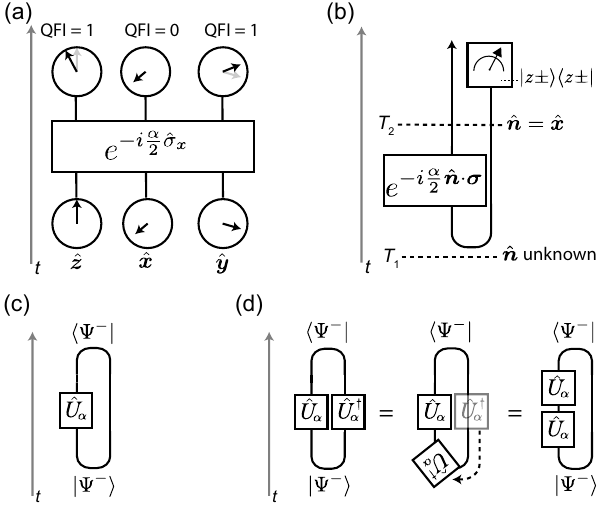}
\caption{{\bf Time-loop metrology:} 
Figures adapted from \cite{24_Song_Agnostic, 25_Song_Superconducting}.
(a) Drawback of sensing the strength of a field oriented in an unknown direction. Time runs vertically. Qubits are prepared in $\hat \sigma_{x,y,z}$ eigenstates. If the field points along $\boldsymbol{\hat{x}}$, the $\hat \sigma_{y,z}$ eigenstates yield the maximum possible QFI per qubit, 1. The $\hat \sigma_x$ eigenstate yields no QFI. On average over the three states, the QFI is $2/3$. 
(b) Hindsight sensing: A probe–ancilla pair begins maximally entangled. The probe evolves under the unknown field, whose direction is then revealed. This information determines how we measure the ancilla to effectively teleport the optimal initial state backward in time to the probe.  
(c) Agnostic sensing: The probe and ancilla begin in a singlet. The probe undergoes the unknown unitary. We then measure whether the qubits remain in a singlet. 
(d) Positronium sensing: The unknown field subjects the qubit to $\hat U_\alpha$ and the antiqubit to $\hat U_\alpha^\dag$, quadrupling the agnostic-sensing QFI.}
    \label{fig:CTC}
\end{figure}

Time-loop metrology circumvents the need for unavailable information at the beginning of conventional metrology experiments. Consider inferring an unknown parameter $\alpha$ encoded in a unitary $\hat U_\alpha = e^{-i\alpha \hat H}$. Ordinarily, to achieve the optimal sensitivity, we must know the forms of eigenstates $\ket{h_\Min}$ and $\ket{h_\Max}$ associated to the Hamiltonian's least and greatest eigenvalues (Sec.~\ref{sec_Intro}). Yet we might lack this information. For example, consider estimating the strength of a magnetic field pointing in an unknown direction $\boldsymbol{\hat {n}}$. The field acts as $\hat U_\alpha = e^{-i\alpha \boldsymbol{\hat {n}} \cdot \boldsymbol{\sigma}/2}$. No single-qubit probe can guarantee information about $\alpha$: we might accidentally prepare the qubit's Bloch vector along $\pm \boldsymbol{\hat {n}}$, preventing the field from affecting the probe \cite{24_Song_Agnostic}. Conventionally, we measure $\alpha$ via quantum-process tomography~\cite{Ariano2001, Altepeter2003}. On average over all possible field directions, quantum-process tomography achieves an FI of $\leq 2/3$ (Fig.~\ref{fig:CTC}a) \cite{24_Song_Agnostic}. 

We can circumvent our ignorance by applying the mathematical equivalence between CTCs and certain quantum circuits. The first such application to metrology, proposed theoretically in ~\cite{23_DRMAS_Non}, enhanced weak-value amplification. Consider measuring a weak coupling's strength, as in Sec.~\ref{sec_wva}. We might learn the optimal initial target state only after coupling target to the probe. Effectively, CTC simulation probabilistically teleports the optimal initial target state backward in time. 

Superconducting-qubit experiments demonstrated two variations on the proposal: hindsight sensing and agnostic sensing~\cite{24_Song_Agnostic}. In \emph{hindsight sensing} (Fig.~\ref{fig:CTC}b), we prepare a probe–ancilla pair in a maximally entangled state (e.g., a singlet). The probe undergoes an unknown unitary $\hat U_\alpha = e^{-i \alpha \hat H}$. Then, we learn the Hamiltonian's form. This information determines the basis in which we measure the ancilla. The measurement effectively teleports the optimal initial state backward in time to the probe. The experiment provided an FI of 0.82; hindsight sensing can achieve a theoretical maximum of 1. The achieved FI exceeded the maximum FI, $2/3$, achievable with two qubits prepared in any separable state. Hence the experiment demonstrated an entanglement advantage despite the entangling gates' finite fidelity. 

Unlike hindsight sensing, \emph{agnostic sensing} never requires knowledge of $\boldsymbol{\hat {n}}$ (Fig.~\ref{fig:CTC}c)~\cite{24_Song_Agnostic}. The aim is to measure the strength $\alpha$ of a field oriented in an unknown direction.  We prepare a probe–ancilla pair in a singlet $\ket{\Psi^-}$ (bottom of Fig.~\ref{fig:CTC}c). The probe undergoes the unitary $\hat U_\alpha = e^{-i \alpha \hat H}$.  We then measure the POVM $\{ \ketbra{\Psi^-}{\Psi^-}, \id - \ketbra{\Psi^-}{\Psi^-} \}$, identifying whether the qubit pair remain in a singlet. If the pair does (top of Fig.~\ref{fig:CTC}c), the experiment simulates a CTC (loop in Fig.~\ref{fig:CTC}c, construed as a spacetime diagram)~\cite{11_Lloyd_Closed}. When the probe undergoes $\hat U_\alpha$, the field perturbs the singlet, reducing the probability of simulating a CTC. This probability varies as $\cos^2(\alpha/2)$, conveying the maximum FI achievable about $\alpha$ with a single-qubit probe, 1. The experiment of~\cite{24_Song_Agnostic} yielded an FI of 0.72, due to the entangling operations' finite fidelity.

Agnostic sensing has been extended in three ways. In one, an experiment introduced \emph{positronium metrology} (Fig.~\ref{fig:CTC}d)~\cite{25_Song_Superconducting}. Positronium is a bound system formed from an electron and a positron. A positron is equivalent to an electron travelling backward in time \cite{42_Stuckelberg,48_Feynman_Relativistic,49_Feynman_Theory,65_Feynman_Nobel}. If a field evolves an electron under $\hat U_\alpha$, it evolves a positron under $\hat U_\alpha^\dag$. Inverting an unknown unitary on a $d$-dimensional system, using a quantum circuit, costs $\mathcal{O}(d^2)$ applications of the unitary~\cite{AmplitudeAmplification, PhaseEstimation, Gily_n_2019}. In contrast, a positron experiences $\hat U_\alpha^\dag$ naturally. Experimentalists implemented similar unitary inversion metrologically, using two transmons~\cite{25_Song_Superconducting}. One served as a qubit; and one, as an \emph{antiqubit}, whose gyromagnetic ratio equaled the negative of the qubit's. The transmons began in a singlet, then evolved under the same unknown field. The qubit underwent $\hat U_\alpha$, whilst the antiqubit underwent $\hat U_\alpha^\dagger$. Finally, the experimentalists measured the singlet-or-not POVM $\{ \ketbra{\Psi^-}{\Psi^-}, \id - \ketbra{\Psi^-}{\Psi^-} \}$. This strategy achieved an FI of 3.02 about $\alpha$ (the theoretical maximum is 4). The experiment motivated a proposal for applying actual antimatter in quantum metrology \cite{25_Wang_Proposal}. 

Agnostic and positronium sensing qualify not only as time-loop metrology, but also as mirror metrology (Sec.~\ref{sec_UUdag}). The initial state $\ket{0}$, in step~\ref{item_prep} of Sec.~\ref{sec_UUdag_QFI}, is a product $\ket{0,0}$ of computational-basis states here. The preparatory unitary $\hat{V}$ there maps $\ket{0,0}$ to $\ket{\Psi^-}$ here. The singlet-or-not measurement here follows from implementing $\hat{V}^\dag$, then measuring whether the state is $\ket{0,0}$ (step~\ref{item_meas} of Sec.~\ref{sec_UUdag}).



In the second adaptation of agnostic sensing, we introduce entanglement not in the preparation and measurement steps, but in the evolution~\cite{Karthik2025}. An ancilla controls whether a single-qubit probe undergoes $\hat U_\alpha$ or $\hat U_\alpha^\dagger$. We prepare the ancilla in a superposition, such that the probe undergoes a superposition of the evolutions. This strategy echos ICO metrology (Sec.~\ref{sec_ICO}), relying on quantum control. However, we do not apply the unitaries with indefinite order here. To infer $\alpha$, one measures the ancilla. This protocol can achieve the optimal FI accessible with a single-qubit probe. However, one might not be able to invert the unknown $\hat{U}_\alpha$, or coherently control it, easily.

In the third extension of agnostic sensing, experimentalists inferred a parameter that characterised a nonunitary channel~\cite{25_Yilmaz_Agnostic}. Let $\boldsymbol{\hat{n}} \cdot \boldsymbol{\sigma}$ denote a Pauli operator dependent on an unknown direction $\boldsymbol{\hat{n}}$. A photon's polarisation underwent dephasing relative to the 
$\boldsymbol{\hat{n}} \cdot \boldsymbol{\sigma}$ eigenbasis. The photon's path degree of freedom served as an ancilla. The experimentalists inferred the dephasing strength agnostically. 

In summary, time-loop metrology demonstrates how mathematical parallels between disparate phenomena -- time travel and quantum sensing -- offer advantages. Hindsight, agnostic, and positronium sensing require no postselection (no discarding of trials)~\cite{24_Song_Agnostic,25_Song_Superconducting,25_Yilmaz_Agnostic}, despite postselection's role in simulating CTCs~\cite{11_Lloyd_Closed}.

Several opportunities merit exploration. First, time-loop metrology may generalise to multiparameter estimation, via states that decouple the target parameter from nuisance parameters. Second, we might agnostically sense nonunitary channels beyond dephasing. Third, the original time-loop-metrology proposal awaits experimental implementation~\cite{23_DRMAS_Non}. We could also extend that proposal from enhancing weak-value amplification to enhancing a variant, partially postselected amplification~\cite{Lupu-Gladstein2021}. 

The antimatter-based proposals, too, suggest opportunities~\cite{25_Song_Superconducting,25_Wang_Proposal}. By extending the strategy in~\cite{25_Song_Superconducting}, we can achieve an estimator variance that scales in Heisenberg fashion with the number of applications of the field. We would subject the qubit–antiqubit pair to the field multiple times sequentially. Additionally, positronium simulation could facilitate quantum algorithms that involve expensive unitary inversion or tomography. Example algorithms include quantum singular-value transformations~\cite{Gily_n_2019,Martyn2021}, amplitude amplification~\cite{AmplitudeAmplification}, and phase estimation~\cite{PhaseEstimation}. Furthermore, many algorithms require the inversion of unitaries representable by $m \times m$ matrices, wherein $m > 2$. We might achieve these inversions by extending positronium metrology from qubits to qudits (multilevel quantum systems) associated with dimensionality-$m$ Hilbert spaces. Ambitiously, we could apply true antimatter to quantum metrology. Near-future technology could realise the proposal in \cite{25_Wang_Proposal}, which may inspire other metrological protocols. 

Finally, we can elaborate on the relationship between time-loop metrology and mirror metrology. Agnostic and positronium metrology qualify as mirror metrology while involving CTC simulations. Do other mirror-metrology experiments involve CTCs or variations thereon? Some mirror-metrology experiments involve bosons, multiparticle entanglement, or submaximal entanglement. In contrast, researchers have argued that qubits, bipartite entanglement, and maximal entanglement can simulate CTCs~\cite{11_Lloyd_Closed}. Can we generalise the theory of CTC simulations to accommodate the latter setups? The effort may illuminate not only connections amongst classes of time-reverse metrology, but also the foundations of CTC theory.

\section{Indefinite-causal-order metrology}
\label{sec_ICO}

\begin{figure}
    \centering
\includegraphics[width=0.5\textwidth]{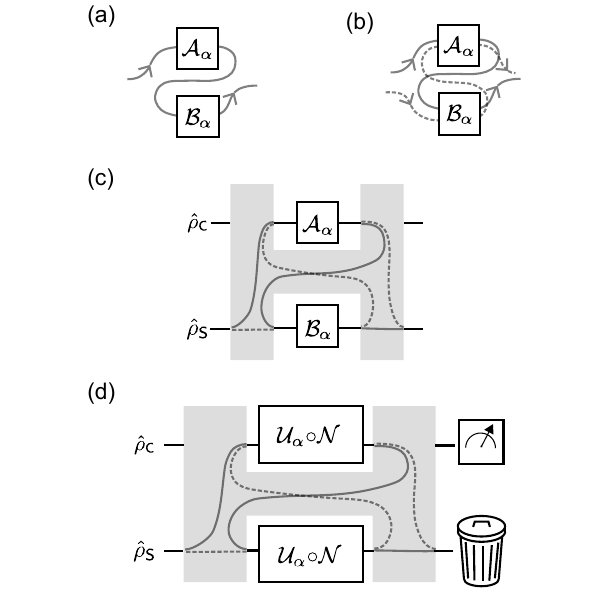}
\caption{{\bf ICO metrology:} 
(a) Channels $\mathcal{A}_\alpha$ and $\mathcal{B}_\alpha$ applied in definite causal order as $\mathcal{B}_\alpha \circ \mathcal{A}_\alpha$. 
(b) ICO process: superposition of $\mathcal{B}_\alpha \circ \mathcal{A}_\alpha$ and $\mathcal{A}_\alpha \circ \mathcal{B}_\alpha$.
(c) The \textsc{switch} (grey H-shaped element) is a supermap that transforms channels $\mathcal{A}_\alpha$ and $\mathcal{B}_\alpha$ on the system $\mathsf S$ of interest. The \textsc{switch} outputs a channel on the control-system composite.
(d) Noise-resilient sensing with readout of just the control: We infer $\alpha$, which parameterises the noisy channel $\mathcal{U}_\alpha \circ \mathcal{N}$, as follows. 
Prepare a control $\mathsf{C}$ in $\hat{\rho}_\mathsf{C}$ and a system $\mathsf{S}$ in  $\hat{\rho}_\mathsf{S}$.
The \textsc{switch} supermap acts on two instances of the channel $\mathcal{U}_\alpha \circ \mathcal{N}$.
$\mathsf{C}$ controls the channel-order superposition.
Measure $\mathsf{C}$, discarding $\mathsf{S}$. The statistics can convey information about $\alpha$ even if the $\mathsf{S}$ readout would be uninformative (e.g., if $\mathsf{S}$ is maximally mixed).}

    \label{fig:ICO}
\end{figure}

Standard quantum circuits apply unitary gates in a fixed sequence. Because each gate causes a state update, the sequence enforces a \textit{definite causal order} (Fig.~\ref{fig:ICO}a).
However, quantum theory effectively allows for indefinite causal order (Fig.~\ref{fig:ICO}b)~\cite{chiribella_beyond_2009, Chiribella_13_Quantum, oreshkov_quantum_2012}. ICO enhances metrological tasks including parameter estimation~\cite{Frey_19_Indefinite, frey_quantum_2021, 21_Chapeau_Blondeau_Noisy, chapeau-blondeau_quantum_2021, chapeau-blondeau_indefinite_2022,chapeau-blondeau_indefinite_2023, 25_Yuan_Indefinite}, channel discrimination~\cite{Chiribella_12_Perfect}, thermometry~\cite{Mukhopadhyay_18_Superposition}, noise mitigation~\cite{Goldberg_23_Evading,Kurdzialek_23_Using}, quantum machine learning~\cite{Gao_23_Measuring}, and geometric-phase measurement~\cite{24_Barnett_Measurement}.
Theoretical proposals have inspired optical experiments~\cite{an_noisy_2024, 23_Yin_Experimental, chen_nonlinear_2025}.
Below, we introduce ICO. We then explain how it can (i) augment the QFI attainable in parameter estimation and (ii) evade noise. We conclude with avenues for future research.

We illustrate ICO with quantum channels, $\mathcal{A}_\alpha$ and $\mathcal{B}_\alpha$, which depend  on an unknown parameter $\alpha$.
Kraus operators
$\hat K_j^{(\mathcal{A}_\alpha)}$ represent $\mathcal{A}_\alpha$: if $\hat \rho$ denotes a quantum state, 
$\mathcal{A}_\alpha(\hat \rho) 
= \sum_{j} \hat K_j^{(\mathcal{A}_\alpha)} \hat \rho 
\big( \hat K_j^{(\mathcal{A}_\alpha)} \big)^\dagger$.
Similarly, $\mathcal{B}_\alpha(\hat \rho)
=\sum_{j} \hat K_j^{(\mathcal{B}_\alpha)} \hat \rho 
\big( \hat K_j^{(\mathcal{B}_\alpha)} \big)^\dagger$. 
If the channels act as $\mathcal{A}_\alpha \circ \mathcal{B}_\alpha$ or $\mathcal{B}_\alpha \circ \mathcal{A}_\alpha$, their order is definite.
In an example of ICO, we apply two channels in a superposition of orders:
the quantum \textsc{switch} is a supermap from a pair of channels, such as ($\mathcal{A}_\alpha, \mathcal{B}_\alpha$), to another channel (Fig.~\ref{fig:ICO}c)~\cite{chiribella_beyond_2009, Chiribella_13_Quantum}.
The output channel acts on a bipartite system: a control qubit $\mathsf{C}$ and a system $\mathsf{S}$ of interest. 
$\mathsf{C}$ controls the superposition enacted by the \textsc{switch}. If $\mathsf{C}$ begins in $|0\rangle$, $\mathsf{S}$ undergoes $\mathcal{A}_\alpha \circ \mathcal{B}_\alpha$.
If $\mathsf{C} $ begins in $|1\rangle$, $\mathsf{S}$ undergoes $\mathcal{B}_\alpha \circ \mathcal{A}_\alpha$.
If $\mathsf{C} $ begins in $|+\rangle \coloneqq (|0\rangle + |1\rangle)/\sqrt{2} \, ,$ the \textsc{switch} epitomises ICO. To show how, we denote $\mathsf{S}$'s initial state by $\hat \rho_\mathsf{S} $.
$\mathsf{CS}$'s final state depends on a forward contribution 
$\hat F_\alpha \coloneqq (\mathcal{A}_\alpha \circ \mathcal{B}_\alpha) (\hat \rho_\mathsf{S} )$,
a reverse contribution 
$\hat R_\alpha\coloneqq (\mathcal{B}_\alpha\circ \mathcal{A}_\alpha)(\hat \rho_\mathsf{S} )$,
and a coherent contribution $\hat C_\alpha \coloneqq \sum_{j,k} \hat K_k^{(\mathcal{B}_\alpha)} 
\hat K_j^{(\mathcal{A}_\alpha)} \hat \rho_{\mathsf{S} } 
\big( \hat K_k^{(\mathcal{B}_\alpha)} \big)^\dagger
\big( \hat K_j^{(\mathcal{A}_\alpha)} \big)^\dagger$: the \textsc{switch} 
induces on $\mathsf{CS}$ the map
\begin{equation}\label{eq:rho-theta-swi}
\begin{aligned}
    |+\rangle\langle+|_\mathsf{C}\!\otimes\!\hat \rho_\mathsf{S}
     \mapsto  \hat \rho_\alpha^{\rm swi}
    \coloneqq  \, \frac{1}{2} &\big(
    \ketbra{0}{0}_{\mathsf{C} }\!\otimes\!\hat F_\alpha 
    + \ketbra{1}{1}_\mathsf{C}\!\otimes\!\hat R_\alpha \\
    &+ \ketbra{0}{1}_\mathsf{C}\!\otimes\!\hat C_\alpha + \ketbra{1}{0}_\mathsf{C}\!\otimes\!\hat C_\alpha^\dagger\big)\,.
\end{aligned}
\end{equation}
The \textsc{switch} implements a type of time reversal by superposing the forward and reverse channel orders.

Debate surrounds experimental realisations of the \textsc{switch}~\cite{Rozema_2024_experimental}.
As an abstract supermap, the \textsc{switch} coherently controls whether $\mathcal{A}_\alpha$ precedes $\mathcal{B}_\alpha$ or vice versa, querying each once~\cite{Chiribella_13_Quantum}. 
One might interpret an experiment as (i) realising this supermap or (ii) simulating the channel~\eqref{eq:rho-theta-swi} using additional platform-specific resources (e.g., extra $\mathcal{A}_\alpha$ or $\mathcal{B}_\alpha$ calls)~\cite{Oreshkov2019timedelocalized,fellous-asiani_comparing_2023,bavaresco_simulating_2025,24_Mothe_Reassessing,Ormrod2023causalstructurein,Vilasini_2024_embedding,MacLean_2017_quantum,Paunkovic2020causalorders}.

To compare definite-causal-order and \textsc{switch}-based strategies, we quantify resources with time (measured relative to the laboratory's rest frame). 
Let $\mathcal{E}_\alpha$ denote the channel parameterised by $\alpha$. Let $\tau$ denote the duration of one platform-level $\mathcal{E}_\alpha$ implementation. 
In metrological contexts, we define a \textit{\textsc{switch} strategy} as any protocol that satisfies two conditions: (i) It induces on $\mathsf{CS}$ the channel~\eqref{eq:rho-theta-swi} wherein $\mathcal{A}_\alpha=\mathcal{B}_\alpha=\mathcal{E}_\alpha$. (ii)  Its continuous evolution (between its preparation and measurement) runs for a duration $\leq 2 \tau$. \textsc{switch} strategies include interferometric realisations. In such an experiment, a control degree of freedom (e.g., a photon's path) coherently routes the probe through the two orders~\cite{Rozema_2024_experimental}. 
Current technologies can implement \textsc{switch} strategies, regardless of the debate.\footnote{
A circuit can simulate a \textsc{switch}~\cite{Chiribella_13_Quantum, bavaresco_simulating_2025}, but in a time $>2 \tau$. The circuit therefore does not implement a \textsc{switch} strategy, according to our definition.}

In Eq.~\eqref{eq:rho-theta-swi}, the coherent contribution $\hat C_\alpha$ enables metrological precision beyond that achievable with definite causal order. 
Early studies focussed on estimating $\alpha$ in an amount $2\tau$ of time.
If $\mathcal{E}_\alpha$ belongs to a certain class of channels, a \textsc{switch} strategy achieves a better measurement precision than
a fixed sequential, definite-causal-order
strategy ($\mathcal{E}_\alpha \circ \mathcal{E}_\alpha$)~\cite{Mukhopadhyay_18_Superposition, Frey_19_Indefinite, frey_quantum_2021, 21_Chapeau_Blondeau_Noisy, chapeau-blondeau_quantum_2021, chapeau-blondeau_indefinite_2022, chapeau-blondeau_indefinite_2023}.\footnote{
Suppose $\mathcal{E}_\alpha$ is nonunitary. No \textsc{switch} strategy is equivalent to $\mathcal{E}_\alpha\circ\mathcal{E}_\alpha$. Although $\hat F_\alpha=\hat R_\alpha$, $\hat C_\alpha\neq \hat F_\alpha$.}$^,$\footnote{
In contrast, \emph{parallel} causally definite strategies can perform as well as \textsc{switch} strategies when applied to specific channels $\mathcal{E}_\alpha$~\cite{24_Mothe_Reassessing}.} 
Examples include thermalising~\cite{Mukhopadhyay_18_Superposition} and depolarising~\cite{Frey_19_Indefinite} channels.
In depolarising-channel estimation, 
we infer the strength $r \in [0, 1]$ of a depolarising channel $\mathcal{N}$. If $\hat{\rho}$ denotes a $d$-dimensional discrete system's state, then $\mathcal{N} (\hat{\rho}) 
= (1-r) \hat{\rho} + r \hat{\id}/d$.
Every \textsc{switch} strategy's QFI exceeds a particular sequential, definite-causal-order strategy's QFI. The relative gain diverges as $r \rightarrow 0$ and vanishes as $r \rightarrow 1$~\cite[Fig.~2]{Frey_19_Indefinite}.

These case studies follow from a general data-processing argument about the output state’s QFI. Let us return to Eq.~\eqref{eq:rho-theta-swi}. Tracing out $\mathsf{C}$ from $\hat \rho_\alpha^{\rm swi}$ yields the mixture $\hat \rho_\alpha^{\rm mix}\coloneqq \tfrac12 \big(\hat F_\alpha+\hat R_\alpha\big)$. 
This tracing out is an $\alpha$-independent completely positive, trace-preserving map. The QFI decreases monotonically under such maps~\cite{ferrie_data-processing_2014}: $\mathcal{I}(\hat \rho_\alpha^{\rm swi})\ge \mathcal{I}(\hat \rho_\alpha^{\rm mix})$. 
We can evaluate $\hat \rho_\alpha^{\rm mix}$ further, assuming $\mathcal{E}_\alpha = \mathcal{A}_\alpha = \mathcal{B}_\alpha$: by the definitions above Eq.~\eqref{eq:rho-theta-swi},
$\hat{R}_\alpha = \hat{F}_\alpha$. Therefore, 
$\hat \rho_\alpha^{\rm mix}
= \tfrac12 \big(\hat F_\alpha+\hat R_\alpha\big)
= \hat{F}_\alpha
= (\mathcal{E}_\alpha\circ \mathcal{E}_\alpha)(\hat \rho_{\mathsf S})$.
This state results from the sequential, definite-causal-order strategy mentioned in the previous paragraph.
Thus, $\mathcal{I}(\hat \rho_\alpha^{\rm swi})\ge \mathcal{I}(\hat F_\alpha)$: every \textsc{switch} strategy achieves at least as much QFI as the sequential, definite-causal-order strategy. 
The \textsc{switch} strategy saturates the bound when the channel’s Kraus operators commute pairwise; noncommutation can yield a quantum advantage.

%

Building on the case studies above, subsequent work sharpened the notion of ICO's metrological advantage. 
We might object to the fairness of pitting a \textsc{switch} strategy against one definite-causal-order baseline. After all, three classes of strategies exist (amongst others): (i) deterministic definite causal order, (ii) classically controlled order, and (iii) quantum-controlled order (e.g., the \textsc{switch})~\cite{liu_optimal_2023, 24_Mothe_Reassessing}. To enable a fair comparison, we should restrict all strategies to the same 
resource budget. Then, we should compare one class's optimal strategy to each other class's. We can bound the greatest QFI attainable from each class, using the process-matrix framework and semidefinite programming~\cite{liu_optimal_2023,24_Mothe_Reassessing}.
ICO improves metrology if class (iii)'s optimal QFI exceeds class (ii)'s, as (ii) forms a superset of (i).

Class (iii) sometimes, but not always, outperforms (ii). Suppose $\mathcal{E}_\alpha$ consists of an $\alpha$-dependent unitary followed by a noise channel. 
Also, suppose the time budget is $2\tau$.
An ICO strategy's QFI exceeds the best definite-causal-order strategies' QFIs~\cite{24_Mothe_Reassessing}.\footnote{
If the time budget is unlimited, however, causal-superposition strategies offer no QFI advantage~\cite{Kurdzialek_23_Using}.}
Now, suppose that $\mathcal{E}_\alpha$ is a qubit-depolarising or -thermalising channel.
The optimal ICO strategy does not outperform the optimal classically controlled strategy~\cite{24_Mothe_Reassessing}. These results refine the earlier narrative: ICO improves metrology only in specific scenarios.

ICO can accomplish more than an increase in QFI: it enables noise-robust sensing (Fig.~\ref{fig:ICO}d). 
Suppose that $\mathcal{E}_\alpha$ unfolds in two steps. First, a channel $U_\alpha$ rotates a qubit through an angle $\alpha$ about an arbitrary known axis. Second, the qubit undergoes a depolarising channel $\mathcal{N}$. Consider inferring $\alpha$ via a \textsc{switch} strategy applied to $\mathcal{E}_\alpha$. 
The depolarisation may leave $\mathsf{S}$ maximally mixed. Nevertheless,
the \textsc{switch}'s control encodes $\alpha$~\cite{Goldberg_23_Evading, 21_Chapeau_Blondeau_Noisy}.
One can estimate $\alpha$ by measuring the control, as demonstrated  in photonic experiments~\cite{23_Yin_Experimental, chen_nonlinear_2025}.
This scheme facilitates channel discrimination~\cite{Chiribella_12_Perfect}, quantum machine learning~\cite{Gao_23_Measuring}, Pancharatnam–Berry-phase measurements~\cite{24_Barnett_Measurement}, and the estimation of products of average displacements~\cite{Zhao_20_Quantum, 23_Yin_Experimental}. The noise resilience extends to multiparameter sensing, if the control corresponds to a Hilbert space of dimensionality $>2$~\cite{Goldberg_23_Evading}. Noise robustness is ICO's most practical metrological gain. It parallels the robustness offered by weak-value amplification against different noise (Sec.~\ref{sec_wva}).

Conceptual and practical questions about ICO metrology remain. First, ICO enables measurements of apparently disparate quantities: unitary parameters~\cite{Mukhopadhyay_18_Superposition, Frey_19_Indefinite, frey_quantum_2021, 21_Chapeau_Blondeau_Noisy, chapeau-blondeau_quantum_2021, chapeau-blondeau_indefinite_2022, chapeau-blondeau_indefinite_2023}, geometric phases~\cite{24_Barnett_Measurement}, quantifiers of operator incompatibility~\cite{Gao_23_Measuring}, etc. We might unify these scattershot applications, systematically characterising when ICO enhances metrology. 

Second, the theory of ICO multiparameter estimation is nascent~\cite{Delgado_2022_symmetries, Goldberg_23_Evading}. Operators' noncommutation can preclude measurements that are optimal for all parameters~\cite{Holevo2011, ragy_compatibility_2016}. We must therefore identify which measurements enable ICO to enhance multiparameter estimation. Further research priorities include (i) establishing a hierarchy of strategy classes~\cite{liu_optimal_2023, 24_Mothe_Reassessing}, (ii) testing whether ICO's advantages persist as the time budget grows~\cite{Kurdzialek_23_Using}, and (iii) proposing platform-specific experiments. 

Third, ICO may facilitate real-world metrology, now that proof-of-principle ICO-metrology experiments have been performed~\cite{23_Yin_Experimental, an_noisy_2024, chen_nonlinear_2025}. We see two main experimental goals: (i) Perform noise-robust parameter estimation by measuring the control system when the sensor becomes maximally mixed~\cite{Goldberg_23_Evading, 21_Chapeau_Blondeau_Noisy}. (ii) Witness
a QFI enhancement over all definite-causal-order strategies, and demonstrate rigorously that the enhancement results from ICO~\cite{procopio_experimental_2015, rubino_experimental_2017, goswami_indefinite_2018, stromberg_demonstration_2023, qu_experimental_2025}.

\section{Outlook}
\label{sec_outlook}

Effective time reversals can enhance quantum metrology. Multiple strategies can effectively reverse time. We have reviewed mirror metrology, weak-value amplification, time-loop metrology, and ICO metrology. Each of Sections~\ref{sec_UUdag}–\ref{sec_ICO} describes research opportunities within the corresponding class of protocols. Here, we describe four opportunities that span, and extend beyond, these classes. 

First, we might unify the classes in an overarching theory of time-reverse metrology. Second, we could cross-pollinate the techniques. For example,~\cite{23_DRMAS_Non} combines CTCs with weak-value amplification. Similarly,~\cite{25_Song_Superconducting,25_Wang_Proposal} combine CTCs with antimatter. Other combinations might merge multiple classes' strengths. Third, antimatter-based sensing~\cite{25_Song_Superconducting,25_Wang_Proposal} has the potential to develop into its own class of time-reverse metrology. 

Fourth, metrology might benefit from effective time-reversal strategies of which the community has not yet conceived. Examples may include the engineering of a low Reynolds number. The Reynolds number characterises fluids, and low values signal laminar flows and negligible inertia~\cite{76_Reynolds_Life}. The corresponding dynamics are reversible. Although the Reynolds number characterises classical systems, quantum analogues have been defined~\cite{13_Jou_Quantum,24_Takeuchi_Quantum}. Hence time reversal holds ample prospects for the future of quantum metrology.

%
%
\begin{acknowledgments}
The authors thank Andrew Jordan, Shimon Kolkowitz, Ana Maria Rey, and Vladan Vuletic for useful conversations.
This work received support from the National Science Foundation under 
QLCI Grant OMA-2120757, 
Grant No.~NSF PHY-1748958, 
and Grant No.~PHY-2408932. 
K.~W.~M. acknowledges funding from the Gordon and Betty Moore Foundation, under grant DOI 10.37807/gbmf11557.
D.~R.~M.~A.~S., K.~W.~M., and N.~Y.~H. thank the Kavli Institute for Theoretical Physics, as well as the organizers of the program ``New Directions in Quantum Metrology,'' for their hospitality. 
Y.-X.W.~acknowledges support from a QuICS Hartree Postdoctoral Fellowship.
\end{acknowledgments}

\begin{appendices}

\onecolumngrid

\renewcommand{\thesection}{\Alph{section}}
\renewcommand{\thesubsection}{\Alph{section} \arabic{subsection}}
\renewcommand{\thesubsubsection}{\Alph{section} \arabic{subsection} \roman{subsubsection}}

\makeatletter\@addtoreset{equation}{section}
\def\theequation{\thesection\arabic{equation}}

\section{Relationship between spin squeezing and bosonic squeezing}
\label{sec:app.spin.squeezing}

Section~\ref{sec_UUdag_squeeze} introduces bosonic squeezing and spin squeezing, techniques used often in mirror metrology. Here, we clarify spin squeezing's relationship with bosonic squeezing, via a spin-to-boson mapping. 

We analyse the following setup. To illustrate spin squeezing, we consider $N \gg 1$ spin-$1/2$ particles. Denote by $\hat S_x$ the global spin angular momentum's $x$-component, and define $\hat S_{y,z}$ analogously. Define $\hat S_-  \coloneqq  \hat S_x - i \hat S_y$ as the global lowering operator. Suppose an external field evolves the spins under $e ^{-i \alpha \hat S _{x} }$. To sense $\alpha$, we initialise the spins in the unit-eigenvalue $\hat S_z$ eigenstate $\ket{0}^{\otimes N}$. 

A spin analogue of the squeeze-probe-unsqueeze protocol amplifies $\alpha$ [Eq.~\eqref{eq:paramp.evol}]. A Holstein-Primakoff transformation elucidates the analogy~\cite{Holstein1940}. Denote by $\hat{a}$ the bosonic-mode lowering operator; and, by 
$\hat Q \coloneqq (\hat a + \hat a^{\dag})/\sqrt{2} \, ,$ the position quadrature operator. The Holstein-Primakoff transformation maps global spin operators to bosonic operators. For example, $\hat S _{-} \mapsto (N -  \hat a ^{\dag}   \hat a  ) ^{\frac{1}{2}}  \hat a $. $\ket{0}^{\otimes N}$ maps to the bosonic vacuum state. If a state contains few excitations ($\langle \hat a ^{\dag}   \hat a \rangle \ll N$), the lowering operator satisfies the approximation 
$\hat S _{-} \approx \sqrt{N} \, \hat a .$ Hence the field acts as $e ^{-i \alpha \hat S _{x} } 
\approx e ^{-i \alpha  \hat Q \sqrt{N}/2 } \, .$
A reverse Holstein-Primakoff transformation maps the bosonic 
$\hat H_\sq (\varphi)$ to a spin-squeezing Hamiltonian $\hat H'_\sq (\varphi) 
\coloneqq \frac{i}{2 N } \, ( e^{-2i \varphi} \, 
\hat S _{-}^2 - \hc )  $. When 
$\varphi \in \{0, \pi/2 \}$, $\hat H'_\sq (\varphi)$ is a \textit{two-axis-twisting} (TAT) Hamiltonian~\cite{Ueda1993}. Not only the TAT Hamiltonian, but also other Hamiltonians can generate spin squeezing. Examples include the one-axis-twisting~\cite{Ueda1993} and Lipkin-Meshkov-Glick~\cite{LMG1965} Hamiltonians.

\section{Derivation of final probe state in weak-value amplification}
\label{sec:app.weak.value.expansion}

Equation~\eqref{eq:weak} shows the state in which the probe ends weak-value amplification. We derive the equation here. References~\cite{Jozsa2007,Dressel&Jordan2012} extend this argument.

First, we review the weak-value-amplification protocol. We prepare the probe in 
$\ket{\Phi_{\rm i}}$ and the target in $\ket{\Psi_{\rm i}}$. Then, we couple the systems via $\hat U_\alpha=e^{-i\alpha \hat A\otimes \hat P}$, measure the target, and postselect on $\ket{\Psi_\mathrm{f}}$. This process yields the (unnormalised) probe state
\begin{align} 
   \ket{\tilde{\Phi}_\mathrm{f}}
   &\coloneqq \bra{\Psi_\mathrm{f}} 
   \hat U_\alpha 
   \left( \ket{\Psi_\mathrm{i}} \ket{\Phi_\mathrm{i}} \right) \\
   \label{Eq_app_phi_f}
   & = \left( \bra{\Psi_\mathrm{f}}  
   \hat U_\alpha
   \ket{\Psi_\mathrm{i}} \right) 
   \ket{\Phi_\mathrm{i}} .
\end{align}

Let us approximate the right-hand side. We Taylor-expand the $\hat U_\alpha$ to first order in  $\alpha$ within the parenthesised factor:
\begin{align}
   \bra{\Psi_\mathrm{f}}\hat U_\alpha\ket{\Psi_\mathrm{i}}
   &=\bra{\Psi_\mathrm{f}}\left[\id-i\alpha \hat A\otimes \hat P
   +\mathcal{O} \left( \alpha^2 \right) \right]
   \ket{\Psi_\mathrm{i}}
   = \braket{\Psi_\mathrm{f}}{\Psi_\mathrm{i}}
   \left[\id-i\alpha A_\mathrm{w}\,\hat P
   +\mathcal{O} \left( \alpha^2 \right) \right] \\
   & = \braket{\Psi_\mathrm{f}}{\Psi_\mathrm{i}}
   \left[ e^{-i \alpha A_{\rm w} \hat{P} }
   + \mathcal{O} \left( \alpha^2 \right) \right] .
   \label{Eq_app_inn_prod}
\end{align}
We have invoked the weak-value definition
$A_\mathrm{w}\coloneqq \bra{\Psi_\mathrm{f}}\hat A\ket{\Psi_\mathrm{i}}/\braket{\Psi_\mathrm{f}}{\Psi_\mathrm{i}}$ [Eq.~\eqref{Eq:wv}].
Let us separate the weak value into its real and imaginary parts:
$A_\mathrm{w}=\mathrm{Re}(A_\mathrm{w})+i\,\mathrm{Im}(A_\mathrm{w})$.
Next, we split the exponential into two exponentials:
\begin{align}
   \bra{\Psi_\mathrm{f}}\hat U_\alpha\ket{\Psi_\mathrm{i}}
   & = \braket{\Psi_\mathrm{f}}{\Psi_\mathrm{i}}
   \left[ e^{-i \alpha {\rm Re} (A_{\rm w}) \hat{P} } \,
   e^{\alpha {\rm Im}(A_{\rm w}) \hat{P} } 
   + \mathcal{O} \left( \alpha^2 \right) \right] .
\end{align}
Having Taylor-approximated the first factor in Eq.~\eqref{Eq_app_phi_f}, we expand the second:
$\ket{ {\Phi}_{\rm i} }
= \int dq \, \phi(q) \ket{q}$.
Equation~\eqref{Eq_app_phi_f} becomes
\begin{align}
   \ket{ \Phi_{\rm f} }
   & = \braket{ \Psi_{\rm f} }{ \Psi_{\rm i} }
   \left[ e^{-i \alpha {\rm Re} (A_{\rm w}) \hat{P} } \,
   e^{\alpha {\rm Im} ( A_{\rm w} ) \hat{P} }
   + \mathcal{O} \left( \alpha^2 \right)
   \right]
   \int dq \, \phi(q) \ket{q} .
\end{align}
We factor out the $e^{i \alpha {\rm Re} (A_{\rm w}) \hat{P} }$ and operate with it on the $\ket{q}$, which becomes $\ket{q + \alpha {\rm Re} (A_{\rm w})}$. Shifting the dummy index $q$ backward, $q \mapsto q - \alpha$, yields
\begin{align} \label{Eq_final_result}
   \ket{ \tilde{\Phi}_{\rm f} }
   & = \braket{ \Psi_{\rm f} }{ \Psi_{\rm i} }
   \left[ e^{\alpha {\rm Im} ( A_{\rm w} ) \hat{P} }
   + \mathcal{O} \left( \alpha^2 \right)
   \right]
   \int dq \,
   \phi \LParen q - \alpha {\rm Re} ( A_{\rm w} ) \RParen
   \ket{q} .
\end{align}
Normalising $\ket{ \tilde{\Phi}_{\rm f} }$ yields the state $\ket{\tilde{\Phi}_{\rm f}}$ in Eq.~\eqref{eq:weak}.
\end{appendices}

%
%
\bibliographystyle{ieeetr}  
\bibliography{Bib_TR_metrology}


\end{document}